\documentclass[12pt]{article}
\input epsf
\newcommand{\NPB}[3]{\emph{ Nucl.~Phys.} \textbf{B#1} (#2) #3}   
\newcommand{\PLB}[3]{\emph{ Phys.~Lett.} \textbf{B#1} (#2) #3}   
\newcommand{\PRD}[3]{\emph{ Phys.~Rev.} \textbf{D#1} (#2) #3}   
\newcommand{\PRL}[3]{\emph{ Phys.~Rev.~Lett.} \textbf{#1} (#2) #3}   
\newcommand{\ZPC}[3]{\emph{ Z.~Phys.} \textbf{C#1} (#2) #3}

\newcommand{\JHEP}[3]{\emph{JHEP} \textbf{#1} (#2) #3}
\def\dalemb#1#2{{\vbox{\hrule height .#2pt
        \hbox{\vrule width.#2pt height#1pt \kern#1pt
                \vrule width.#2pt}
        \hrule height.#2pt}}}

 \def\bd{\begin{document}} \def\ed{\end{document}}
\def\ds{\documentstyle} \let\fr=\frac \let\bl=\bigl \let\br=\bigr
\let\Br=\Bigr \let\Bl=\Bigl 
\let\bm=\bibitem
\let\na=\nabla
\let\pa=\partial \let\ov=\overline
\def\ie{{\it i.e.\ }} 
\newcommand{\pr}{\paragraph{}}
\newcommand{\be}{\begin{equation}}
\newcommand{\ee}{\end{equation}}
\newcommand{\beba}{\begin{equation}\begin{array}{lcl}}
\newcommand{\eaee}{\end{array}\end{equation}}
\newcommand{\bea}{\begin{eqnarray}}
\newcommand{\eea}{\end{eqnarray}}
\newcommand{\ba}{\begin{array}}
\newcommand{\ea}{\end{array}}
\newcommand{\td}{\tilde}
\newcommand{\norsl}{\normalsize\sl}
\newcommand{\ns}{\normalsize}
\newcommand{\refs}[1]{(\ref{#1})}
\def\simlt{\mathrel{\lower2.5pt\vbox{\lineskip=0pt\baselineskip=0pt
           \hbox{$<$}\hbox{$\sim$}}}}
\def\simgt{\mathrel{\lower2.5pt\vbox{\lineskip=0pt\baselineskip=0pt
           \hbox{$>$}\hbox{$\sim$}}}}

\begin{document}
\thispagestyle{empty}
\rightline{\normalsize\sf hep-ph/0004091}
\rightline{\normalsize CERN-TH/2000-103}
\rightline{\normalsize CPHT-S013.0300}
\rightline{\normalsize IEM-FT-201/00}
\rightline{\normalsize LPTENS-00/16}
\rightline{\normalsize April 2000}
\vskip 1.0truecm
\centerline{\Large\bf 
Radiative symmetry breaking in brane models}
\vskip 1.truecm
\centerline{{\large\bf I. Antoniadis}~$^a$, 
{\large\bf K. Benakli~$^b$} and {\large\bf M. Quir{\'o}s}~$^{c,d}$}
\vskip .5truecm
\centerline{{\it $^a$Centre de Physique Th{\'e}orique, Ecole Polytechnique,
91128 Palaiseau, France}
\footnote{Unit{\'e} mixte du CNRS et de l'EP, UMR 7644.}}
\vskip .5truecm
\centerline{{\it $^b$CERN Theory Division
 CH-1211, Gen{\`e}ve 23, Switzerland}}
\vskip .5truecm
\centerline{{\it $^c$Laboratoire de Physique Th{\' e}orique, ENS, 24 rue
Lhomond, F-75231, Paris, France}
\footnote{Unit{\'e} mixte du CNRS et de l'ENS, UMR 8549.}}
\centerline{\it $^d$Instituto de Estructura de la Materia (CSIC),
Serrano 123, E-28006 Madrid, Spain}

\vskip 0.5truecm
\centerline{\bf\small ABSTRACT}
\vskip .4truecm

We propose a way to generate the electroweak symmetry breaking radiatively in
non-supersymmetric type I models with string scale in the TeV region. By
identifying the Higgs field with a tree-level massless open string
state, we find that a negative squared mass term can be generated at one loop. 
It is finite, computable and typically a loop factor smaller than the
string scale, that acts as an ultraviolet cutoff in the effective
field theory. When the Higgs open string has both ends confined on our 
world brane, its mass is predicted to be around 120 GeV, i.e. that of
the lightest Higgs in the minimal supersymmetric model for large 
$\tan\beta$ and $m_A$. Moreover, the string scale turns out to be one
to two orders of magnitude higher than the weak scale. We also discuss 
possible effects of higher order string threshold corrections that
might increase the string scale and the Higgs mass.

\hfill\break
\vfill\eject

Following the recent understanding of string theory, the string
scale, $M_s$, is not tied to the Planck mass but corresponds to 
an independent arbitrary parameter~\cite{Ant}-\cite{st}, restricted by present
experimental  data to be $M_s\simgt 1$ TeV~\cite{lowgrav}. Therefore
a non-supersymmetric string model with
a string scale in the TeV range provides a natural solution,
alternative to supersymmetry, to the
gauge hierarchy problem~\cite{add,ab}. For such models 
an important question is to understand the origin of
electroweak symmetry breaking, and explain
the mild hierarchy between the weak and string scales. In string models
all tree-level masses are fixed by the string scale, except for flat
directions that give arbitrary masses to the fields that couple to
them. This implies that electroweak symmetry breaking should occur
radiatively in two possible ways: a) If the Higgs corresponds to a
massless field with a quartic tree-level potential, and a negative squared
mass is generated by string one-loop radiative corrections which
are not protected by supersymmetry. b) If the Higgs vacuum
expectation value (VEV) is classically undetermined by a flat
direction which is lifted radiatively and fixed at a local minimum
of the effective potential.

In this Letter we study these issues in the context of 
type I string models possessing non-supersymmetric brane
configurations~\cite{ads,au}. We will first present a one-loop computation of
the effective potential in the presence of a Wilson-line background that
corresponds to a classically flat direction. We will show that the
resulting potential has a non-trivial minimum which fixes the 
VEV of the Wilson line or, equivalently, the
distance between the branes in the $T$-dual picture. Although the
obtained VEV is of the order of the string scale, the potential
provides a negative squared-mass term when expanded around the
origin. 
Next we discuss models, obtained 
by orbifolding the previous example, where the Wilson line is projected 
away from the spectrum while keeping charged massless fields with
quartic tree-level terms. These fields acquire one-loop negative
squared masses, that can be computed using the previous calculation.
By identifying them with the Higgs field we can achieve radiative
electroweak symmetry breaking~\footnote{For an earlier attempt to
generate a non-trivial
minimum of the potential, see Ref.~\cite{gg}.}, and obtain the mild hierarchy
between the weak and string scales in terms of a loop factor. 

This mechanism becomes very predictive in a class of models where the
Higgs field corresponds to a charged massless excitation of an open string
with both ends confined on our world brane (analog to the untwisted states of 
heterotic orbifolds). In this case, the tree-level potential can be
obtained by an appropriate tree-level truncation of a supersymmetric
theory leading to two predictions. On the one hand, the Higgs mass
is predicted to be that of the lightest Higgs in the minimal
supersymmetric model (MSSM) for large values of $\tan\beta$ and $m_A$,
i.e. $\sim$ 120 GeV~\cite{higgs}. On the other hand, 
the string scale is computable and turns out to be around one to two orders of
magnitude higher than the weak scale, roughly $1-10$ TeV. This mechanism is
similar to the Coleman-Weinberg idea, except that there are no logarithms in
the computation. Indeed, from the field theory point of view the string scale
provides an ultraviolet cutoff which regulates the quadratic
divergence of the Higgs mass. Finally, we discuss higher order string
threshold corrections which can affect the above results, for instance
by large logarithms when there are massless bulk fields that propagate in two
large transverse dimensions~\cite{ab,abd}. In this case, the string scale and
possibly the Higgs mass could be pushed up to higher values.

The reader who is not familiar with string theory
could skip the following rather technical section and go directly to
Eq.~(\ref{mu2R}) and Fig.~\ref{eps}, 
which provides an estimate of the generated
string one-loop mass term for a tree-level massless scalar on our
world brane.

\section*{One-loop effective potential}

Here we will consider a simple non-supersymmetric tachyon-free $Z_2$ orientifold 
of type IIB superstring compactified to four dimensions on
$T^4/Z_2\times T^2$~\cite{ads}. Cancellation of Ramond-Ramond charges
requires the presence of 32 D9 and 32 anti-D5 (D$\bar5$)
branes~\footnote{In general arbitrary numbers of pairs D9+D$\bar9$ and    
D5+D$\bar5$ can also be added~\cite{au}.}. The bulk (closed strings) 
as well as the D9 branes are
$N=2$ supersymmetric while supersymmetry is broken on the world-volume of the
D$\bar5$'s. The massless closed string spectrum contains
the graviton-, 19 vector- and 4
hyper-multiplets, while the massless open string spectrum 
on the D9 branes contains an $N=2$ vector multiplet
in the adjoint of the $SO(16)\times SO(16)$ gauge group and a
hypermultiplet in the {\bf (16,16)} representation. When all D$\bar5$
branes are put at the origin of $T^4$, the
non-supersymmetric D$\bar5$ sector contains gauge fields and 
complex scalars in the adjoint representation of $USp(16)\times
USp(16)$ gauge group, a pair of complex scalars in the {\bf (16,16)}
representation, and Dirac fermions in the {\bf (120,1)} + {\bf (1,120)} +
{\bf (16,16)} representations. Finally there are 9$\bar5$ strings giving rise
to complex scalars in the {\bf (16,1;1,16)} + {\bf (1,16;16,1)} together with
Weyl fermions in the {\bf (16,1;16,1)} + {\bf (1,16;1,16)}
representations, with respect to $SO(16)\times SO(16)\times
USp(16)\times USp(16)$. Note that the $9\bar 5$ spectrum is supersymmetric when
D$\bar 5$ gauge interactions are turned off.

We will restrict ourselves to the effective potential involving the
scalars of the D$\bar5$ branes, namely the adjoints and bifundamentals
of the $USp(16)\times USp(16)$ gauge group. The relevant part of the
one-loop partition function corresponding to $\bar5\bar5$ open strings
is
\begin{eqnarray}
\label{particion}
{\cal A}_{\bar5}&=&\frac{1}{4}(d_1+d_2)^2\frac{V_8-S_8}{\eta^8}
W_4 P_2+\frac{1}{4}
(d_1-d_2)^2\frac{V_4 O_4-O_4 V_4-C_4 C_4+S_4 S_4}{\eta^8}\left( 
\frac{2 \eta}{\theta_2} \right)^2 P_2
\nonumber\\
{\cal M}_{\bar5}&=&\frac{1}{4}(d_1+d_2)\left\{
\frac{\widehat{V}_8+\widehat{S}_8}{\widehat\eta^8}W_4 
+\frac{\widehat V_4 \widehat O_4-\widehat O_4 \widehat V_4+
\widehat C_4 \widehat C_4-\widehat S_4 \widehat S_4}{\widehat\eta^8}\left( 
\frac{2 \widehat\eta}{\widehat\theta_2} \right)^2 \right\}P_2
\end{eqnarray}
where ${\cal A}$ and ${\cal M}$ denote the contributions from the
annulus and M{\"o}bius strip, respectively. In the above equation $d_1=d_2=16$,
while $V_{2n}$, $O_{2n}$, $C_{2n}$ and $S_{2n}$ 
are the $SO(2n)$ characters,
$$
V_{2n}=\frac{\theta_3^n-\theta_4^n}{2\eta^{n}}\, , \quad
O_{2n}=\frac{\theta_3^n+\theta_4^n}{2\eta^{n}}\, , \quad
C_{2n}=\frac{\theta_2^n-i ^n \theta_1^n}{2\eta^{n}}\, , \quad
S_{2n}=\frac{\theta_2^n+i ^n \theta_1^n}{2\eta^{n}}\, , 
$$ 
where $\theta_i$ are the Jacobi theta functions and
$\eta$ the Dedekind eta function, depending on the usual complex variable
$\tau=it/2$, with $t$ being the (real) annulus parameter. In the product of
characters, the first factor stands for the contribution of 
space-time and $T^2$ world-sheet fermions, while the second factor
represents the corresponding contribution from the internal $T^4$. 
The hatted functions are
defined by $\widehat f\equiv f(\tau+1/2)$. Finally, $P_2$ ($W_4$) denotes 
the momentum (winding) lattice sum along the $T_2$
($T_4$) torus~; for one dimension, they read:
\be
P_1(\tau)=\sum_m e^{2i\pi\tau m^2\alpha'/ R_\parallel^2}\qquad ;\qquad
W_1(\tau)=\sum_n e^{2i\pi\tau n^2 R_\perp^2/\alpha'}\, ,
\label{lat}
\ee
where $\alpha'\equiv M_s^{-2}$ is the Regge slope, and $R_\parallel$
($R_\perp$) denotes the radius of the corresponding dimension parallel
(transverse) to the D-brane.

In both the annulus and  M{\"o}bius amplitudes the first term stands for
the  untwisted contribution while the second term accounts for the 
$Z_2$ orbifold projection which differentiates $T^4$ and $T^2$
contributions. Its presence is due to the non-freely action of $Z_2$
at the origin of $T^4$ and thus it depends only on the lattice of $T^2$.
It is obvious from Eq.~(\ref{particion}) that the $Z_2$ projection 
acts in a supersymmetric way, and therefore the second terms containing
the twisted contribution vanish identically and will not play any role in our
calculation.

In the first terms containing the untwisted contribution, 
$V_8$ and $S_8$ arise from bosons and fermions,
respectively. Here, supersymmetry is explicitly broken via the orientifold
projection realized by the M{\"o}bius amplitude. Indeed, from the
change of sign of $S_8$ between $\cal A$ and $\cal M$, it is manifest
that the orientifold projection acts in opposite ways for bosons and 
fermions and breaks supersymmetry. More precisely, it symmetrizes the
bosons and antisymmetrizes the fermions in each $USp(16)$ factor.

The tree-level scalar potential can be obtained by a truncation of an $N=2$
supersymmetric theory and has flat directions corresponding to the Wilson
lines $a$ along the $T^2$ or $T^4$ directions. For longitudinal directions
they amount to shifting the momenta $m\to m+a$ in Eq.~(\ref{lat}), while for
transverse directions they shift the windings $n\to n+a$ and describe 
brane separation. It follows that at one-loop level the flat directions are
lifted since the Wilson lines acquire a potential from the M{\"o}bius
amplitude which breaks supersymmetry. Without loss of generality
we will consider a Wilson line $a$ along one direction of $T^2$
of radius $R$, and treat the other, upon T-duality, on the same footing as
the dimensions of $T^4$ with a common radius $r$. After transforming the
amplitudes (\ref{particion}) in the transverse (closed string) channel
and using the standard
$\theta$-function Riemann identity, the one loop effective potential for the
Wilson line is given by~:
\bea
\label{Veff}
V_{\rm eff}(a) &=& {1\over 32\pi^4\alpha'^2}\int_0^\infty dl
{\theta_2^4\over 4\eta^{12}}\left(il+{1\over 2}\right) {R\over r^5}
\sum_{\vec m}e^{-2\pi{{\vec m}^2\over r^2}l} \sum_n e^{-4i\pi na}
e^{-2\pi n^2R^2l}\\
&=& {1\over 32\pi^4\alpha'^2}\int_0^\infty dl
{\theta_2^4\over 4\eta^{12}}\left(il+{1\over 2}\right) {R\over r^5}
\sum_{\vec m}e^{-2\pi{{\vec m}^2\over r^2}l}\left(
1+2\sum_{n>0}\cos(4\pi na)e^{-2\pi n^2R^2l}\right)\, ,\nonumber
\eea
where the radii $R$ and $r$ are defined in units of $\alpha'$.

In this setup, the canonically normalized scalar field $h$ associated to the
Wilson line $a$ is $h=a/gR$, where $g$ is the gauge coupling, as can be easily
seen by dimensional reduction. Let us first expand the effective potential in
powers of $h$ and extract its quadratic (squared mass) term $\mu^2 h^2/2$. The
result is:
\be
\mu^2 =-{g^2\over 2\pi^2\alpha'}\int_0^\infty dl
{\theta_2^4\over 4\eta^{12}}\left(il+{1\over 2}\right) {R^3\over r^5}
\sum_{\vec m}e^{-2\pi{{\vec m}^2\over r^2}l}\sum_n n^2 e^{-2\pi
  n^2R^2l}\ .
\label{mu2}
\ee
It is easy to see that the integral converges. In fact, in the limit
$l\to\infty$ the integrand falls off exponentially, while for $l\to 0$ one
can use the Poisson resummations
\bea
\sum_m e^{-2\pi{m^2\over r^2}l} &=& {r\over\sqrt{2l}}
\sum_p e^{-\pi{r^2\over 2l}p^2}\, ,
\label{poisson}\\
\sum_n n^2 e^{-2\pi n^2R^2l} &=& {1\over R\sqrt{2l}}\sum_n\left(
{1\over 4\pi R^2l}-{n^2\over 4R^4l^2}\right) e^{-{\pi\over 2R^2l}n^2}\, ,
\label{poissonR}
\eea
and the identity
$$
{\theta_2^4\over \eta^{12}}\left(il+{1\over 2}\right)=(2l)^4
{\theta_2^4\over \eta^{12}}\left({i\over 4l}+{1\over 2}\right)\, ,
$$
to show that the integrand goes to a constant. Moreover, $\mu^2$ is negative
which implies that the origin is unstable and $h$ must acquire a non trivial
VEV breaking the gauge symmetry. Note that the negative sign comes
from the expansion of $\cos 4\pi n a$ in Eq.~(\ref{Veff}) and is
correlated with the positive sign of the contribution from the same states to
the cosmological constant. Although this seems to be a general
property in these models, we do not have a deeper understanding of the
correlation between the sign of the mass term and the (massive)
spectrum of the theory. 

Even if $a$ is a periodic variable of period 1, $V_{\rm eff}$ is
periodic under the shift $a\to a+1/2$, since its contribution originates from
the M{\"o}bius amplitude. Moreover, in this particular example, the one-loop
effective potential has a global minimum at $a=1/4$. This follows trivially
from its expression (\ref{Veff}), whose derivative with respect to $a$ is a
sum of terms proportional to $\sin 4\pi na$, while its second derivative gives
\be
V''_{\rm eff}\mid_{a=1/4} = {1\over 2\pi^2\alpha'^2}\int_0^\infty
dl {\theta_2^4\over 4\eta^{12}}\left(il+{1\over 2}\right) {R\over r^5}
\sum_{\vec m}e^{-2\pi{{\vec m}^2\over r^2}l} \sum_n (-)^{n+1}n^2
e^{-2\pi n^2R^2l}\, .
\label{Veffmass}
\ee
Positivity of the integrand is manifest for all factors with the
exception of the last sum for which a careful analysis is required.
This sum can be written as $\partial_\tau\theta_4(\tau)/2i\pi$ with 
$\tau=2\,iR^2l$, which can be easily shown to be a positive function.

In the $T$-dual picture, the VEV $a=1/4$ corresponds to separating a brane at
a distance from the origin equal to half the compactification interval
$\pi R$. By
turning on all Wilson lines $a_I$, the effective potential becomes a sum
$\sum_I V_{\rm eff}(a_I)$, with $V_{\rm eff}(a_I)$ given in (\ref{Veff}),
which upon minimization fixes all $a_I$ at the same value $1/4$. Thus, the
global minimum of all Wilson lines corresponds to put all branes at the same
point in the middle of the compactification interval. The $USp(16)\times
USp(16)$ gauge group is then broken down to a $U(8)\times U(8)$ or $USp(16)$
subgroup, corresponding to turning on Wilson lines along the $T^2$
or $T^4$ directions, transforming in the adjoint or in the bifundamental
representation, respectively.

\begin{figure}[htb]
\centering
\epsfxsize=4.5in
\epsffile{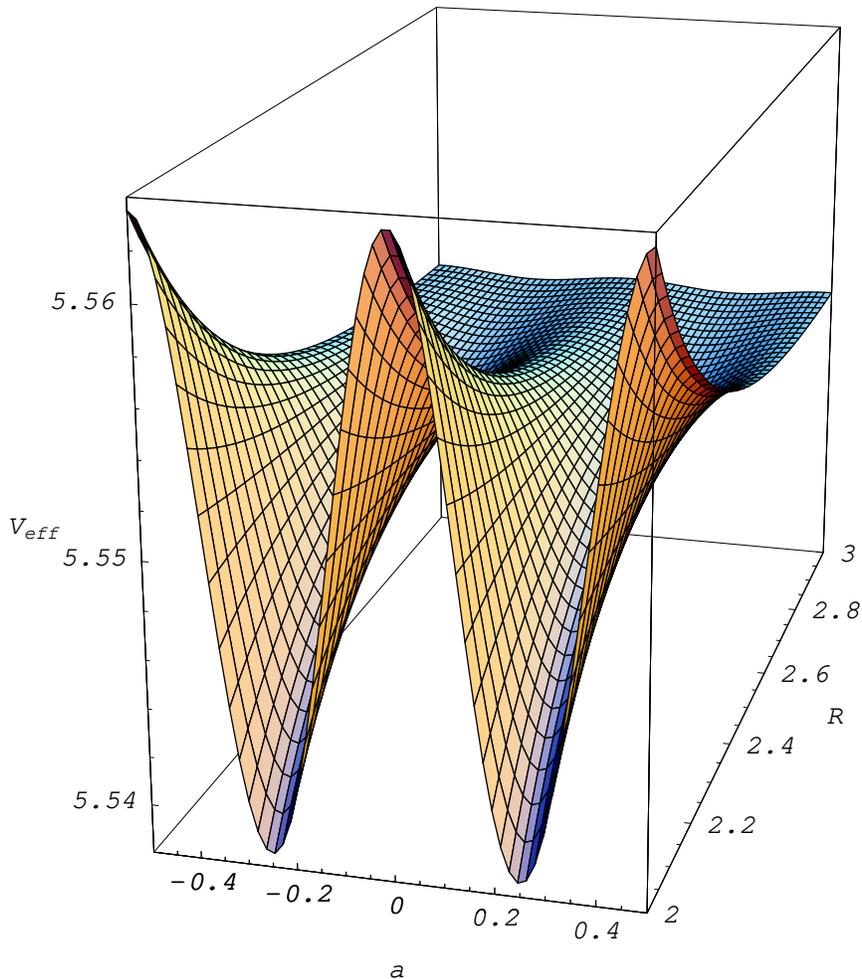}
\caption{Effective potential as a function of $a$, for $2\leq R\leq 3$
in units of $10^{-4}M_s^4$.}
\label{effpot}
\end{figure}

In order to make a numerical estimate of the results, we will consider the case
of a 4-brane with five large transverse dimensions by taking the limit
$r\to\infty$ and keeping the radius $R$ (along the 4-brane) as a
parameter. To take the limit $r\to\infty$, we use Eq.~(\ref{poisson})
for each of the five transverse dimensions, and note that only $p=0$
contributes in the sum. In fact, non vanishing values
of $p$ may contribute only in the region $l\to\infty$, in which case the
corresponding integrand in Eq.~(\ref{Veff}) vanishes as $l^{-5/2}$.
It follows that in the limit $r\to \infty$ the potential becomes:
\be
V_{\rm eff}(a,R) = {R\over 32\pi^4\alpha'^2}\int_0^\infty 
\frac{dl}{\left(2\, l\right)^{5/2}}
{\theta_2^4\over 4\eta^{12}}\left(il+{1\over 2}\right) \sum_n e^{-4i\pi na}
e^{-2\pi n^2R^2l}\, .
\label{Veff1}
\ee
The effective potential (\ref{Veff1}) is plotted in Fig.~\ref{effpot} for the
range of values of the radius $2\leq R\leq 3$ as a function of 
$a$ inside its period $(-1/2, 1/2)$. Following our previous analysis, 
it has a maximum at the origin and a minimum at $a=\pm 1/4$ for any
value of $R$.

\begin{figure}[htb]
\centering
\epsfxsize=4.in
\epsffile{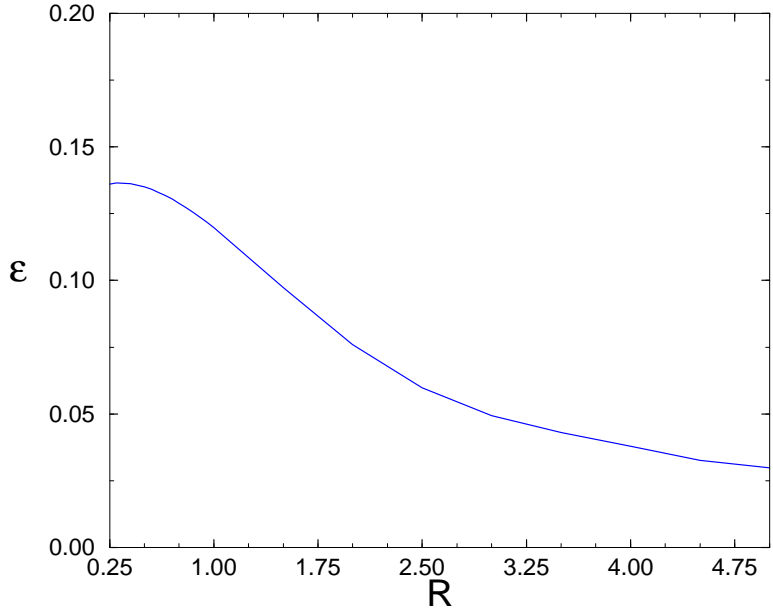}
\caption{The parameter $\varepsilon$ in (\ref{epsilon2R}) as a
function of $R$ in $\alpha'$ units.}
\label{eps}
\end{figure}

The mass term at the origin, in the limit $r\to\infty$ and for arbitrary $R$,
can be equally computed from Eq.~(\ref{mu2}) using the Poisson
resummation (\ref{poisson}). The result is:
\be
\label{mu2R}
\mu^2(R)=-\varepsilon^2(R)\, g^2\, M_s^2
\ee
with
\be
\varepsilon^2(R) ={1\over 2\pi^2}\int_0^\infty \frac{dl}{\left(2\,
l\right)^{5/2}} 
{\theta_2^4\over 4\eta^{12}}\left(il+{1\over 2}\right) R^3
\sum_n n^2 e^{-2\pi n^2R^2l}\ .
\label{epsilon2R}
\ee
The parameter $\varepsilon$ is plotted in Fig.~\ref{eps} as a function of $R$
in a typical range $1/4<R<5$. 
At the lower end, it has almost reached its
asymptotic value for $R\to 0$~\footnote{This limit corresponds, upon T-duality,
to a large transverse dimension of radius $1/R$.},
$\varepsilon(0)\simeq 0.14$, and the effective cutoff for the mass
term at the origin is $M_s$, as can be seen from Eq.~(\ref{mu2R}). At
large $R$, $\mu^2(R)$ falls off as $1/R^2$, which is the effective
cutoff in the limit $R\to\infty$, in agreement with field theory
results in the presence of a compactified extra
dimension~\cite{SS}~\footnote{Actually this effect is at the origin of thermal
squared masses, $\sim T^2$, in four-dimensional field theory at finite
temperature, $T$, where the time coordinate is compactified on a circle
of inverse radius $1/R\equiv T$ and the Boltzmann suppression factor
generates an effective cutoff at momenta $p \sim T$.}. In fact, in the
limit $R\to\infty$ an analytic approximation to $\varepsilon(R)$ can be
computed as,
\be
\varepsilon(R)\simeq \frac{\varepsilon_\infty}{M_s\, R}\, ,
\qquad\qquad
\varepsilon_\infty^2=\frac{3\, \zeta(5)}{4\, \pi^4}\simeq 0.008\, ,
\label{largeR}
\ee
which approximately describes Fig.~\ref{eps} for large values of $R$.

Notice that the mass term (\ref{mu2R}) we found for the Wilson line $a$
also applies, by gauge invariance, to the charged massless fields
which belong to the same representation.

\section*{Electroweak symmetry breaking}

In the previous example we obtained a VEV of the order of the string
scale, because we only considered Wilson lines, which correspond
to tree-level flat directions in the Cartan subalgebra of the gauge
group, and have put to zero the VEV's of all other fields. Thus, the total
potential to be minimized appeared at the one-loop level. Had we minimized
the effective potential with respect to fields charged under the Cartan
subalgebra, we would have found the same solution
(which corresponds to a true minimum in the multidimensional field space)
since the charged fields acquire, from the Wilson lines, positive tree-level
squared masses and have vanishing VEV's. In more realistic models, the Wilson
lines are at least partially projected away by an orbifold projection which
also breaks the gauge group. If the orbifold projection acts in a
supersymmetric way, as was the case of the $Z_2$ in the previous example, the
calculation of the squared mass term remains valid for the left-over charged
scalars in the spectrum, up to an overall numerical factor given by the order
of the orbifold group ($1/N$ for a $Z_N$ orbifold). Moreover, the charged
scalars have a tree-level potential which can be obtained by an appropriate
truncation, dictated by the orbifold, of a supersymmetric theory. These two
facts allow the existence of a (local) perturbative minimum, around which
higher order terms in the expansion of the one loop potential can be neglected
since the charged scalars would acquire a VEV controlled by the quadratic
terms.

We will illustrate these points within the context of the toy model described
in the previous section. The crucial property  is that the bosonic sector of
the non-supersymmetric D$\bar 5$ branes is identical to the one of an 
$N=2$ supersymmetric theory  obtained by a
$Z_2$ orbifold projection from an $N=4$ 
theory based on a ``fictitious" $USp(32)$ gauge group. The latter
contains six adjoint scalars that can be organized in three $N=1$
chiral multiplets $\Phi_i$ with $i=1,2,3$. 
Notice that in this model supersymmetry is explicitly broken because
the fermions belong to the antisymmetric instead of the adjoint
(symmetric) representation of $USp(32)$.
The  $Z_2$ 
projection breaks  
$USp(32)$ into $USp(16) \times USp(16)$ and keeps the adjoint of
$USp(16) \times USp(16)$ 
from $\Phi_1$ and the $\bf{(16,16)}$ components from  $\Phi_{2,3}$.
The tree-level scalar potential can be obtained straightforwardly 
by a corresponding truncation of the potential of the $N=4$ theory: 
\be
V_{N=4}=\frac {g^2}{2}{\rm Tr}\left(\sum_{i,j}\left| \left[ \Phi_i, \Phi_j 
\right] \right|^2 + \left(
\sum_{i}  \left[ \Phi_i, \Phi_i^{\dagger} \right]\right)^2 \right)\, .
\label{n4potential}
\ee
The result is identical to the potential of an $N=2$ theory with 
$USp(16) \times USp(16)$ gauge group and one hypermultiplet in the 
$\bf{(16,16)}$ representation. In $N=1$ notation, it corresponds to the
superpotential $W=g/\sqrt{2}\ \varphi_2 \varphi_1 \varphi_3$ where
$\varphi_1$ is the adjoint from $\Phi_1$ and $\varphi_{2,3}$ are the two
bifundamental chiral multiplets from $\Phi_{2,3}$. 
The $F$- and $D$-term contributions
to the potential come from the first and second term of
Eq.~(\ref{n4potential}), respectively.  

As we discussed in detail after  Eq.~(\ref{lat}), the $Z_2$ orbifold projection
does not by itself break all supersymmetries and does not play any role in the
computation of the potential. As a result, the scalar mass terms generated at
one loop receive contributions only from the untwisted sector which treats the
adjoint and the $\bf{(16,16)}$ scalars in the same way, as an adjoint of
$USp(32)$. Thus, the generated masses of the different scalars can be obtained
from the same functional of the radii through permutations. In particular, this
means that scalars describing displacement of branes in dimensions of the same
size acquire equal masses. For instance, in the isotropic 3-brane limit of six
large transverse dimensions, $r\to\infty$ and $R\to 0$, the result
(\ref{mu2R}) applies for all scalar components.

We would like now to discuss possible phenomenological applications of these
results. Let us assume that there is a sequence of ``supersymmetric"
orbifold projections that lead to the Standard Model living on some
non-supersymmetric brane configuration along the line of the toy model
presented above. In the minimal case, where there is only one Higgs
doublet $h$ originating from the untwisted sector, the scalar potential would
be:
\be
V=\lambda (h^\dagger h)^2 + \mu^2 (h^\dagger h)\, ,
\label{potencialh}
\ee
where $\lambda$ arises at tree-level and is given by an appropriate truncation
of a supersymmetric theory. Within the minimal spectrum of the Standard Model,
$\lambda=(g_2^2+g'^2)/8$, with $g_2$ and $g'$ the $SU(2)$ and $U(1)_Y$ gauge
couplings, as in the MSSM. On the other hand,
$\mu^2$ is generated at one loop and can be estimated by
Eqs.~(\ref{mu2R}) and (\ref{epsilon2R}).

The potential (\ref{potencialh}) has a minimum at $\langle h\rangle
=(0,v/\sqrt{2})$, where $v$ is the VEV of the neutral component of the $h$
doublet, fixed by $v^2=-\mu^2/\lambda$. Using the relation of $v$ with the $Z$
gauge boson mass, $M_Z^2=(g_2^2+g'^2)v^2/4$, and the fact that the quartic
Higgs interaction is provided by the gauge couplings as in supersymmetric
theories, one obtains for the Higgs mass a prediction which is the
MSSM value for $\tan\beta\to\infty$ and $m_A\to\infty$:
\be
\label{masa}
M_h=M_Z\ .
\ee
Furthermore, one can compute $M_h$ in terms of the string scale
$M_s$, as $M_h^2=-2\mu ^2=2\varepsilon^2 g^2M_s^2$, or equivalently
\begin{equation}
M_s=\frac{M_h}{\sqrt{2}\, g\varepsilon}
\label{final}
\end{equation}

The lowest order relations (\ref{masa}) and (\ref{final}) receive in
general two kinds of higher order corrections. On the one hand, there
might be important string corrections that we will discuss in the next
section. On the other hand, from the point of view of the effective
field theory, they are valid at the string scale $M_s$, and
Standard Model radiative corrections
should be taken into account for scales between $M_s$ and $M_Z$. 
In particular, the tree level Higgs mass has been shown to receive important
radiative corrections from the top-quark sector. For present experimental
values of the top-quark mass, the Higgs mass in Eqs.~(\ref{masa}) and
(\ref{final}) is raised to values around 120 GeV~\cite{higgs}. Moreover from
Eq.~(\ref{final}), we can compute the string scale $M_s$. There is a first
ambiguity in the value of the gauge coupling $g$ at $M_s$, which depends on
the details of the model. Here, we use a typical unification value $g\simeq
1/\sqrt{2}$. A second ambiguity concerns the numerical coefficient
$\varepsilon$ which is in general model dependent. In our calculation, this is
partly reflected in its $R$-dependence, as seen in Fig.~\ref{eps}. Varying $R$
from 0 to 5, that covers the whole range of values for a transverse dimension
$1<1/R<\infty$, as well as a reasonable range for a longitudinal dimension
$1<R\simlt 5$, one obtains $M_s\simeq 1-5$ TeV. 
Note that in the $R\gg 1$ (large longitudinal dimension) region our
theory is effectively cutoff by $1/R$ and the Higgs mass is then
related to it by, 
\begin{equation}
\frac{1}{R}=\frac{M_h}{\sqrt{2}\, g\,\varepsilon_\infty}\, .
\label{finalR}
\end{equation}
Using now the value for $\varepsilon_\infty$ in the present model,
Eq.~(\ref{largeR}), we find $1/R\simgt 1$ TeV.

A further model dependence of
$\varepsilon$ comes from the order of the orbifold group. As mentioned above,
had we considered a higher order orbifold, e.g. $Z_{2N}$ instead of $Z_2$ as
required by more realistic models, $\varepsilon$ would decrease by a factor
$\sqrt N$. As a result, the radiative electroweak symmetry breaking can be
consistent with a string scale as heavy as ${\cal O}(10)$ TeV and a
compactification scale $1/R\simgt 2$ TeV.

In a more general context, the Higgs sector may be more complicated
and the scalar potential could have classically undetermined flat
directions as discussed in the introduction. For concreteness we will
consider the case of two Higgs doublets $h_1$ and $h_2$ with a tree-level
potential, obtained by an appropriate truncation of a
supersymmetric theory, and equal to that of the MSSM. We are also
assuming two different one-loop generated squared mass terms $\mu_1^2$ and
$\mu_2^2$ for the Higgs fields:
\be
\label{potencial2}
V=\lambda \left(\left|h_1\right|^2-\left|h_2\right|^2\right)^2
+\rho \left|h_1^* h_2\right|^2 +\mu_1^2 \left|h_1
\right|^2+\mu_2^2 \left|h_2 \right|^2 
\ee
where $\lambda=(g_2^2+g'^2)/8$ and $\rho=g_2^2/2$.
The conditions for having a stable minimum are
$\mu_1^2+\mu_2^2>0$ and $\mu_1^2 \mu_2^2<0$. These conditions are
fulfilled provided that one of the masses, say $\mu_2^2$, is negative
and the other, say $\mu_1^2$, is positive. In this case we get the VEV's
$\langle h_1 \rangle =0$ and $\langle h_2 \rangle =(0,v/\sqrt{2})$, where 
$v^2=-\mu_2^2/\lambda$. Using again the relation of $v$ with $M_Z$, we
obtain the tree-level Higgs mass spectrum~:
\be
M_{h_2}=M_Z\, , \qquad M_{h_1^0}^2=\mu_1^2+\mu_2^2\, ,
\qquad M_{h_1^-}^2=M_{h_1^0}^2+M_W^2\, ,
\label{espectro}
\ee
where $h_2$ corresponds to the Standard Model Higgs, and $h_1^0$, $h_1^-$ to
the neutral and charged components of the $h_1$ doublet. Moreover, the string
scale is given by
\begin{equation}
M_s=\frac{M_{h_2}}{\sqrt{2}\, g\varepsilon_2}
\label{final2}
\end{equation}
with $\mu_2^2=-\varepsilon_2^2 g^2M_s^2$. 

Again, these are tree-level relations which are subject to both string and
Standard Model radiative corrections. In particular, the latter
provide important contributions to the mass of the Standard Model Higgs $h_2$, 
which is increased roughly to $\sim$ 120 GeV, and accordingly to the string
scale given in Eq.~(\ref{final2}). It is interesting that we obtained the same
relations as in the previous example with a single Higgs field. The difference
is that there is also a left-over scalar doublet whose neutral
and charged components acquire masses given in Eq.~(\ref{espectro}). 
As we have pointed out, in this case one needs  the one-loop
generated squared masses for the two scalar doublets, $\mu_1^2$, $\mu_2^2$, to
be different and  opposite in sign. Although our toy string example allows for
different values by introducing different radii, the change in sign
requires more general models, such as those obtained for instance by
introducing additional pairs of branes - anti-branes~\cite{au}.

\section*{Discussion on string threshold corrections}

We discuss now string threshold corrections to the relations (\ref{masa})
and (\ref{final}). These are moduli dependent and may
become very important only when some radii become large compared to the string
length. Otherwise, if all radii are of order one in string units, higher loop
corrections are order one numbers multiplied by loop factors which are
suppressed when string theory is weakly coupled. Of course, these (model
dependent) corrections are needed for a detailed phenomenological analysis and
could be as important as those of the MSSM that increase the Higgs mass by
roughly 10\%. An estimate of these corrections can be done by an explicit
computation of the $a^4$ terms in the expansion of the potential (\ref{Veff1}).
Notice though that these terms do not determine uniquely the one-loop
corrections to the quartic couplings of the charged fields, partly because there
are more than one gauge invariant combinations. An additional subtlety is the
existence of an infrared divergence as $l\to 0$, which is due to the low energy
running of the couplings and must be appropriately subtracted to obtain the
string threshold corrections in a definite renormalization scheme~\cite{aq}.

For dimensions longitudinal to our world brane, the large radius limit leads in
general the theory very rapidly to a non perturbative regime, since the
(ten-dimensional) string coupling becomes strong when four-dimensional
gauge couplings are of order unity. On the other hand, for large
transverse dimensions, the tree-level string coupling remains perturbative (of
order of the gauge couplings), and therefore their size can in principle become
as large as desired. If this is the case, the decompactification limit exists,
and threshold corrections are again controlled by the string coupling and are
suppressed by loop factors. However, this limit does not exist in general when
there are massless bulk fields that propagate in one or two transverse
dimensions, and threshold corrections become very important~\cite{ab}.

A way to see how these large corrections to the parameters of the effective
lagrangian on the brane arise, is to look at the ultraviolet open string loop
diagrams as emission of massless closed strings in the bulk at the
location of distant sources created by other branes or orientifold planes.
This emission leads to corrections that diverge linearly or logarithmically
with the size of transverse space, if there are massless closed string states
propagating in one or two dimensions, respectively. The case of one large
transverse dimension is similar to that of a large longitudinal
one, since threshold corrections grow linearly with the radius and bring
rapidly the theory to a non-perturbative regime~\cite{pw}. In this case, one
can fine-tune the radius to a narrow region near the string scale and the low
energy parameters will be very sensitive to the initial conditions. 

In the case of two large transverse dimensions, the logarithmic contributions
to the parameters of the effective action on the brane are similar to those in
a renormalizable theory and can be resummed as in the renormalization group
improved MSSM~\cite{ab}. In this analogy, the string scale $M_s$ plays the
role of the supersymmetry breaking scale, while the size of the transverse
space replaces the ultraviolet cutoff at the Planck mass, $M_P$. For instance,
if the bulk contains $n$ large transverse dimensions of common radius 
$R_\perp$, while the remaining $6-n$ have string size, one obtains
the familiar relation $M_P^2=M_s^{2+n} R_\perp^n$. When there are
massless bulk fields propagating in two of them, like e.g. twisted moduli
localized at an $n-2$ dimensional subspace, the logarithmic
corrections are $\propto \log(R_\perp M_s)=(2/n)\log(M_P/M_s)$. 

Concerning the Higgs mass considered here, such large radius dependent
contributions would arise if there are bulk massless
fields emitted by the Higgs at zero external momentum. The vanishing of such
tree-level couplings, as for instance with bulk gravitons, implies the absence
of large threshold corrections for the Higgs mass at the one-loop level. This
is in agreement with our result (\ref{mu2}) which remains finite in the
decompactification limit for any number of large transverse
dimensions. However, large corrections can arise at higher orders, e.g. 
through gravitons emitted from open string loops. While computation of such
effects is out of the scope of this work, we would like to discuss the general
structure of such corrections and comment on their phenomenological
implications.

In the simplest case, the relevant part of the world brane action in the
string frame is:
\be
{\cal L}_{\rm brane}=e^{-\phi}\left\{\omega^2|DH|^2+{1+\tan^2\theta_W\over 8}
\omega^4(H^\dagger H)^2 +{1\over 4}(F_{SU(2)}^2+\cot^2\theta_W F_Y^2)
\right\}-\varepsilon^2M_s^2\omega^4|H|^2\, ,
\label{Lbrane}
\ee
where $\phi$ is the string dilaton, $\omega$ the scale factor of the
four-dimensional (world brane) metric, $H$ the Higgs scalar (in the string
frame) and $D$ the gauge covariant derivative. The weak angle at the string
scale $\theta_W$ must be correctly determined in the string model.
Notice that the last term has no $e^\phi$ dependence since it corresponds to a
one loop correction. The bulk fields $\phi$ and $\omega$ are evaluated in the
transverse coordinates at the position of the brane. The physical couplings
$g_2$, $\lambda$ and the mass $\mu^2$ are given by
\be
g_2=e^{\phi/2}\ ,\qquad
\lambda={1+\tan^2\theta_W\over 8}\, e^\phi\ ,\qquad 
\mu^2=-\varepsilon^2e^\phi\omega^2M_s^2\, ,
\label{param}
\ee
while Eq.~(\ref{masa}) remains unchanged and the relation (\ref{final}) becomes
\be
\label{finalc}
M_s=\frac{M_h}{\sqrt{2}\, \varepsilon\, e^{\phi/2}\omega}\, .
\ee
The lowest order result (\ref{final}) corresponds to the (bare) value $\omega=1$.

As we discussed above, when the bulk fields $\phi$ and $\omega$ propagate in
two large transverse dimensions, they acquire a logarithmic dependence on
these coordinates due to distant sources. Since the value of $\phi$ at the
position of the world brane is fixed by the value of the gauge coupling in
Eq.~(\ref{param}), the relation (\ref{masa}) for the Higgs mass is not
affected, while Eq.~(\ref{finalc}) for the string scale is corrected by a
renormalization of $\omega$ which takes the generic form:
\be
\omega=1+b_\omega g_2^2\ln (R_\perp M_s)\, ,
\label{omega}
\ee
where $b_\omega$ is a numerical coefficient. This correction is similar to a
usual renormalization factor in field theory, which here is due to an infrared
running in the transverse space. Depending on the sign of $b_\omega$, it can
enhance ($b_\omega<0$) or decrease ($b_\omega>0$) the value of the string scale
by the factor $1/\omega$. This effect can be important since the involved
logarithm is large, varying between 7 and 35, for $R_\perp$ between 1 fm and 1
mm.

In more general models, there are additional bulk fields entering in the
expression of low energy couplings on the brane, such as the twisted moduli
localized at the orbifold fixed points. As a result, every term in the
lagrangian (\ref{Lbrane}) may be multiplied by a different combination of the
bulk fields that acquires an independent correction, similarly to
Eq.~(\ref{omega}). Thus, in the generic case, both relations (\ref{masa}) and
(\ref{final}) may be modified by corresponding renormalization factors that are
computable in every specific model. In particular,
the prediction of $\sim 120$ GeV for the Higgs mass, which coincides
with that of the lightest Higgs in the MSSM for large values of
$\tan\beta$ and $m_A$, can change by this effect.

A final important question that we have not addressed in this letter
is the possible signatures of Higgs production in brane world
models. Previous works done in the context of the effective field
theory suggest that there may be new effects, leading in general to
signatures that are different from those in the Standard Model or the
MSSM~\cite{higgss}. It will be interesting to study this issue in the
framework of the non-supersymmetric type I string models we discussed
here.    

\section*{Acknowledgments}

This work was partly supported by the EU under TMR contracts 
ERBFMRX-CT96-0090 and ERBFMRX-CT96-0045, by
CICYT (Spain) under contract  AEN98-0816 and by IN2P3-CICYT contract Pth 96-3.
K.B. thanks the CPHT of Ecole Polytechnique for hospitality.


\end{document}